\documentclass[12pt,titlepage]{utarticle}

\usepackage{amsmath,amssymb,amsbsy,amstext, afterpage}
\usepackage{graphicx}
\usepackage{amsfonts}
\usepackage{amssymb}
\usepackage{float}

\numberwithin{equation}{section}
\newcommand{\be}{\begin{equation}}
\newcommand{\bea}{\begin{eqnarray}}

\newcommand{\ee}{\end{equation}}
\newcommand{\eea}{\end{eqnarray}}
\newcommand{\ba}{\begin{array}}
\newcommand{\ea}{\end{array}}

\newcommand{\nc}{\newcommand}
\nc{\gYM}{g_{\mathrm{YM}}}
\nc{\geff}{g_{\mathrm{eff}}}
\nc{\qIR}{q_{\mathrm{IR}}}

\begin{document}

\baselineskip=15.5pt

\title{Constraining holographic inflation \\[1ex] with WMAP}

\author{\hspace{-0.3cm}Richard Easther$^1$, Raphael Flauger\address{
      Department of Physics,\\
      Yale University,\\
      New Haven, CT 06520, USA\\
   }, 
      Paul McFadden$^2$ $\&$ Kostas Skenderis\address{
      Institute for Theoretical Physics,\\
   $^3$ KdV Institute for Mathematics, \\
$^4$ Gravitation and Astro-Particle Physics Amsterdam, \\ 
      Science Park 904,
      1090 GL Amsterdam, the Netherlands\\
      {~}\\[.5cm]
      \emailt{\tt richard.easther@yale.edu, }
      \emailt{\tt raphael.flauger@yale.edu, }
      \emailt{\tt p.l.mcfadden@uva.nl, }
      \emailt{\tt k.skenderis@uva.nl }
   }$^{\,3\,4}$
}

\Abstract{
In a class of recently proposed models, the early universe is strongly coupled and described holographically by a three-dimensional, weakly coupled, super-renormalizable quantum field theory. This scenario leads to a  power spectrum of scalar perturbations that differs from the usual empirical $\Lambda$CDM form and the predictions of generic models of single field, slow roll inflation.  This spectrum is characterized by two parameters: an amplitude, and a parameter $g$ related to the coupling constant of the dual theory.
We estimate these parameters, using WMAP and other astrophysical data.
We compute Bayesian evidence for both the holographic model and standard $\Lambda$CDM and find that their difference is not  significant, although $\Lambda$CDM provides a somewhat better fit to the data.   However, it appears that Planck will permit a definitive test of this holographic scenario.
   \vspace{0.1cm}
}

\maketitle

\tableofcontents
\bigskip

\section{Introduction}

It was recently proposed that the very early universe is in a strongly coupled and non-geometric phase,  best described holographically as a
weakly coupled three-dimensional quantum field theory in flat spacetime with no gravity  \cite{McFadden:2009fg,McFadden:2010na,McFadden:2010jw,McFadden:2010vh}.   In this scenario, the strongly coupled phase plays the same role as inflation in more conventional models of the very early universe.  The corresponding predictions for the scalar and tensor power spectra and non-Gaussianity were worked out in \cite{McFadden:2010na} and \cite{McFadden:2010vh} (see \cite{McFadden:2010jw} for a summary).  In the present paper,  we confront these models with the observational data from WMAP and other sources, and compare them to the fit provided by the standard 
power-law $\Lambda$CDM model.

This holographic scenario is qualitatively different from conventional models of inflation. The latter are
described using graviton and inflaton fluctuations about a homogeneous and isotropic inflating background.
In the new holographic models, however, the very early universe is  such that notions of spacetime, and
perturbations around it, are not yet well-defined. Rather, they  emerge at the end of the strongly coupled epoch, after which the universe is described by usual  hot big bang cosmology. Such a non-geometric period should have a well-defined description in string theory in terms of a strongly coupled sigma model; here, we use holography to model it in terms of a weakly coupled three-dimensional QFT.  Conventional inflationary models can also be described using the holographic framework,  but these scenarios are dual to strongly coupled QFTs.

A wealth of  observational data now allows us to test theoretical ideas about the early universe, providing a probe of fundamental physics at energies close to the Planck scale.  Power-law $\Lambda$CDM, an empirical model with only six free parameters, fits the existing data remarkably well.\footnote{Cosmological fits to CMB data usually include a parameter quantifying the contribution of the Sunyaev-Zel'dovich effect to the apparent small-scale temperature anisotropies. We marginalize over this parameter, but will not discuss it further in what follows.}  This model describes a flat universe with radiation, baryons, cold dark matter, a cosmological constant and a power-law spectrum of adiabatic primordial fluctuations. Four of the six parameters describe the composition and expansion of the universe, namely the Hubble rate $H_0=100 h\,\mathrm{km/s/Mpc}$, the physical baryon and dark matter densities $\Omega_b h^2$ and $\Omega_c h^2$, and the optical depth due to re-ionization $\tau$.  Given that we do not need  spatial curvature to fit the data, the current dark energy contribution  follows from the requirement that the overall density of the universe is equal to the critical value.  The remaining two parameters characterize the  power spectrum of primordial curvature perturbations
\be \label{power-law}
\Delta_\mathcal{R}^2(q) = \Delta_\mathcal{R}^2(q_*) \left(\frac{q}{q_*}\right)^{n_s-1},
\ee
where $\Delta_\mathcal{R}^2(q_*)$ is the amplitude and   $n_s$ the spectral tilt.  The measured value of  $\Delta_\mathcal{R}^2(q_*)$ depends on an arbitrary reference scale, or pivot, $q_*$.

This is a purely empirical parameterization but there is no  evidence that the  primordial power spectrum  is not well-described by this choice (see {\em e.g.\/} \cite{Peiris:2009wp}). Moreover, while almost all slow roll inflationary models  predict  that $n_s$ is  a function of $q$,  the dependence is typically weak. This ``running'' is expressed in terms of $\alpha_s = d n_s/d\ln{q}$, evaluated at $q_*$.  In simple inflationary models, $\alpha_s$ is higher order in slow roll than  $n_s-1$, the departure from scale invariance \cite{Kosowsky:1995aa}.

Interestingly, the holographic model predicts  a scalar power spectrum of the form
\be \label{hol_power}
\Delta_\mathcal{R}^2(q) =\Delta_\mathcal{R}^2 \frac{1}{1+ (g q_*/q) \ln |q/gq_*|},   
\ee
where $g$ is a free parameter that replaces $n_s$. This power spectrum corresponds to a 2-loop approximation on the QFT side, and higher loop terms would modify this formula.  Roughly speaking, 
$g q_*/q$ corresponds to the effective coupling constant, and for self-consistency it should be small for all cosmologically relevant values of $q$.
When this is the case, one can transform (\ref{hol_power}) into the form of (\ref{power-law}) with a spectral index $n_s=n_s(q)$ that is now scale-dependent \cite{McFadden:2010na}. In this case, however, all logarithmic derivatives $d^k n_s(q)/d \ln q^k$ are of the same order, in contrast to the slow roll case, and the expansion cannot be truncated.
A principal goal of this paper is to determine whether a spectrum of the form (\ref{hol_power}) is consistent with the data and to quantify its ability to explain the observed universe, relative to that of the conventional power spectrum. In doing so, we are contrasting the predictions of holographic cosmology with both the empirical $\Lambda$CDM  model and the large class of slow roll inflationary scenarios whose power spectrum is effectively described by (\ref{power-law}).    The parameter set for ``holographic $\Lambda$CDM'' is ($\Omega_b h^2$, $\Omega_c h^2$, $h$, $\tau$, $\Delta_\mathcal{R}^2$, $g$), while the $\Lambda$CDM  parameter set takes its usual form.

The form of  (\ref{hol_power}) follows uniquely from the basic properties of a certain class of dual QFTs.  If the data are incompatible with (\ref{hol_power}) we could then exclude this entire class of holographic models. More precisely, the dual model is a super-renormalizable QFT admitting a 't Hooft large-$N$  limit,
with massless fields and a single dimensionful coupling constant. The momentum dependence of (\ref{hol_power}) essentially follows  from dimensional analysis: the effective coupling of the theory is obtained by forming the unique dimensionless combination of the dimensionful coupling constant and the momentum $q$.
As mentioned above, (\ref{hol_power}) corresponds to a 2-loop approximation, and all momentum dependence enters solely through the dependence
of this computation on the effective coupling. In particular, the scale-invariance of  (\ref{hol_power}) at large $q$ is directly related to the asymptotic freedom of super-renormalizable theories.

Given that the  computation of (\ref{hol_power}) rests on two approximations, the large-$N$ limit and a perturbative 2-loop approximation, in addition to asking whether  holographic $\Lambda$CDM  fits the data, we must also check that the best-fit values of  $\Delta_\mathcal{R}^2$ and $g$ are compatible {\em a posteriori\/} with these theoretical approximations. The smallness of the amplitude $\Delta_\mathcal{R}^2$ immediately implies that $N \sim 10^4$, so the large-$N$ approximation is indeed valid.  Moreover, the best-fit value of $g$ is very small, reflecting the  near scale-invariance of the power spectrum, but it will be important to check that the effective coupling is small for all values of $q$  which  contribute to the observed CMB anisotropies and are thus constrained by the WMAP data.

Our data analysis is performed using a suitably modified version of CAMB and CosmoMC~\cite{Lewis:2002ah,cosmomc}, together with the MultiNest sampler~\cite{Feroz:2007kg,Feroz:2008xx}.  As well as speeding the parameter estimations relative to the usual MCMC sampler\footnote{The performance of an MCMC sampler is a function of the allowed parameter ranges, stepsizes, data and the underlying model. We could not find a set of parameters for the usual MCMC sampler with the holographic model for which the acceptance rate was close to that typically found for $\Lambda$CDM, resulting in chains that converged only very slowly.}, this combination of tools  efficiently computes Bayesian evidence.  Evidence  discriminates between two or more competing models, and is sensitive to the likelihood across the overall parameter space of each model  \cite{JaynesBK,Trotta:2008qt}. In addition, we compare the maximum likelihood values found for each model, and the corresponding information criteria.

This paper is organized as follows.  In Section 2, we summarize the theoretical calculation of the power spectrum for holographic $\Lambda$CDM.
Section 3 examines the parameter estimates obtained for this model and the appropriate priors for the free parameters.   Section 4  discusses the model selection problem, and Section 5 contains a discussion of our results.

As this paper was being completed \cite{Dias:2011in} appeared, which also presents  parameter constraints for the holographic $\Lambda$CDM power spectrum. However, the parameter estimation problem is approached somewhat differently here, and we also address   model selection.

\section{Theoretical prediction} \label{sec:theo-pre}

In this section we describe the theoretical prediction for the
power spectrum of the holographic model.

In general, a holographic model is specified by providing the dual
three-dimensional QFT and the holographic
formulae that relate the cosmological observables of interest to
correlation functions of the dual QFT. There are two classes of
three-dimensional theories that one may currently use to model
the very early universe: the first class consists of three-dimensional
QFTs with a non-trivial fixed point in the UV, while the second
class involves super-renormalizable QFTs. The models we
discuss here belong to this second class. We emphasize that the
theoretical predictions discussed below are specific to this class
of models.

More precisely then, a theory in this second class should have the following properties: (i)
it should admit a large-$N$ limit, (ii) all fields should be massless,
(iii) it should have a dimensionful coupling constant, (iv) all terms
in the Lagrangian should have the same scaling dimension, which should
be different from three. Properties (ii)-(iv) imply that the theory
admits a generalized conformal structure
\cite{Jevicki:1998ub,Kanitscheider:2008kd}, meaning that the theory
would be conformal if the coupling constant is promoted to a
background field that transforms under conformal transformations. A
class of models exhibiting these properties is given by three-dimensional
$SU(N)$ Yang-Mills theory coupled to a number of massless
scalars and fermions, all transforming in the adjoint of $SU(N)$,
with interactions consisting of Yukawa terms and quartic scalar
terms. In three dimensions, the Yang-Mills coupling constant
$\gYM^2$ has dimension one, and for the models we discuss,
one may arrange (by rescaling the fields appropriately) that the
coupling constant appears only as an overall constant in the action.
Assigning scaling dimension one to scalars and gauge fields, and
$3/2$ to fermions, one finds that kinetic terms and the interactions have
dimension four. Theories of this type appear as the
worldvolume theories of D-branes.

In our case, we are interested in the primordial power spectrum of scalar perturbations:
this may be obtained from the 2-point function for the trace of the stress
tensor $T$ as we discuss below. The generalized conformal structure and large-$N$
counting imply that the general form of the
2-point function at large $N$ is\footnote{If one  \label{all-order}
imposes only the generalized conformal structure then the r.h.s.~of
(\ref{2pt}) is modified as
$ N^2 f(\geff^2) \to f(N^2\hspace{-2pt},\, \geff^2)$, where $f(N^2\hspace{-2pt},\, \geff^2)$ is a general
function of two variables.} \cite{Kanitscheider:2008kd},
\be
\label{2pt}
\langle T(q)T(-q) \rangle =  q^3 N^2 f(\geff^2),
\ee
where $q$ is the magnitude of the 3-momentum, $\geff^2=\gYM^2 N/q$ is the
effective dimensionless 't Hooft coupling and $ f(\geff^2)$ is
a general function of  $\geff^2$.

\begin{figure}[t]
\center
\includegraphics[width=5.6cm]{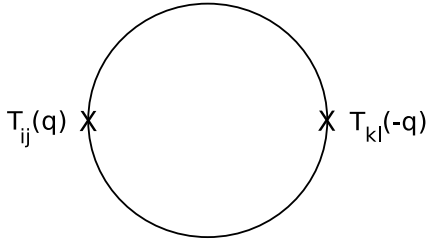}
\hspace{2pc}
\begin{minipage}[b]{14pc}
\caption{\label{1-loop-dia}
1-loop contribution to $\langle T(q) T(-q) \rangle$.  We sum over the contributions from gauge fields, scalars and fermions, with each diagram yielding a contribution of order $\sim N^2 q^3$.}
\end{minipage}
\end{figure}

The holographic framework of \cite{McFadden:2009fg} involves
the analytic continuation
\be \label{anal-cont}
q \to - i q, \qquad N \to -i N,
\ee
and the formula for scalar power spectrum is
\be \label{holo_formula}
\Delta_\mathcal{R}^2(q) = - \frac{q^3}{4 \pi^2}
\frac{1}{{\rm Im} \langle T(q) T(-q)\rangle} ,
\ee
where the imaginary part is taken after applying the analytic continuation.
Note that under (\ref{anal-cont}) we have
\be
\geff^2 \to \geff^2, \qquad N^2 q^3 \to - i N^2 q^3.
\ee
It follows that in theories with generalized conformal invariance,
the 2-point function of the stress tensor transforms very simply
under the analytic continuation (\ref{anal-cont}). We thus
obtain the final formula for the scalar power spectrum:
\begin{equation} \label{holo_formula-f}
\Delta_\mathcal{R}^2(q) = \frac{1}{4 \pi^2 N^2} \frac{1}{f(\geff^2)}.
\end{equation}
Including subleading $1/N^2$ corrections to this formula is
straightforward (one just expands the function $f(N^2\hspace{-2pt},\, \geff^2)$ in $1/N^2$, see footnote \ref{all-order}), but, as we shall see,
the data favor $N \sim 10^4$ so such terms are in practice negligible.\footnote{
On the other hand, it is unclear
how to incorporate effects that are non-perturbative in $N$: exponentially
small corrections of the type $\exp(- N^2)$ are mapped to exponentially
large effects after the analytic continuation. In the discussion above it is
understood that the large-$N$ limit is taken before the analytic
continuation is performed.}

\begin{figure}[t]
\center
\includegraphics[width=0.8\textwidth]{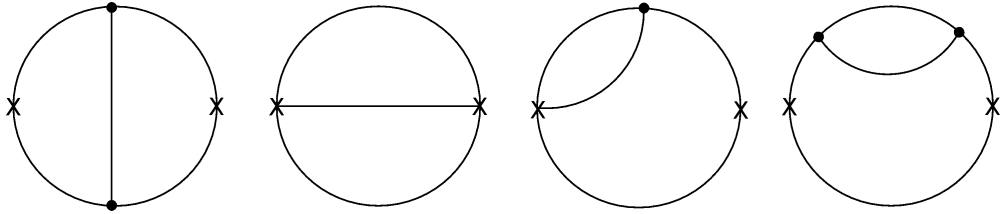}
\caption{\label{2-loop-fig} Diagram topologies contributing at 2-loop order.}
\end{figure}

When $\geff^2$ is small, one may compute the function $f(\geff^2)$ and one finds 
that it has the form
\be \label{f}
f(\geff^2) = f_0(1 - f_1 \geff^2 \ln \geff^2 + f_2 \geff^2 + O[\geff^4]).
\ee
The leading-order term $f_0$ is determined at 1-loop in perturbation theory
(see Fig.~\ref{1-loop-dia}), and its precise value may be found in \cite{McFadden:2010na}.
The constant $f_1$ is determined by a standard 2-loop computation. 
As is well known, in perturbation theory super-renormalizable theories with massless fields display severe infrared divergences.
Indeed, each of the 2-loop diagrams listed in Fig.~\ref{2-loop-fig} evaluates to
an overall factor of $N^3 g_{\mathrm{YM}}^2$ multiplying an integral with
superficial degree of (infrared) divergence two.  Imposing an infrared cut-off, $\qIR$, one may evaluate the integrals to obtain $\sim q^2 \ln (q/\qIR)$. Altogether, one finds a 2-loop contribution to the stress tensor 2-point function of the order
\be
\label{QFTresult}
N^2 q^3 g_{\mathrm{eff}}^2 \ln (q/\qIR) = N^2 q^3 (-\geff^2 \ln \geff^2 + \geff^2 \ln (\gYM^2 N/\qIR)).
\ee
Thus, $f_1$ is determined but $f_2$ is still undetermined since so far $\qIR$ is arbitrary.
It was argued in \cite{Jackiw:1980kv}, however, that this infrared divergence is an artefact of perturbation theory and the theory develops a physical scale that acts as a cut-off. In~\cite{Tom}, it was shown for a specific class of models that
a large-$N$ resummation indeed leads to a finite answer with $\qIR \sim \gYM^2 N$.
Having thus obtained the infrared scale, $f_2$ is then determined unambiguously.

By rearrangement, (\ref{f}) may also be written as
\be
\label{f_expansion}
f(\geff^2) = f_0(1 + f_1 \geff^2 \ln (1/(f_3 \geff^2)) + O[\geff^4]),
\ee
where $f_3 = \exp (-f_2/f_1)$. The constants $f_1, f_2, f_3$ in general depend on all parameters of the
theory ({\it i.e.}, the field content, Yukawa and quartic couplings), but have not been computed to date.
As long as we probe the theory at scales far above the infrared scale $\qIR$, however, the specific value of $f_3$
should only provide a small correction since $|\ln\geff^2|\gg|\ln f_3|$. We will thus write $f_3 =\beta |f_1|$ and take $\beta=1$ in the following;
later, in Section \ref{sec:fudge}, we will check how the fit to the data changes if $\beta$ is allowed to vary.
To simplify our notation, we set
\be
\label{g_def}
 f_1 \gYM^2 N = g q_*\, ,
\ee
where $q_*$ is the pivot scale, taken here to be $q_*=0.05$ $\mathrm{Mpc}^{-1}$.
Substituting back into (\ref{holo_formula-f}), we finally obtain the power spectrum written down in equation~(\ref{hol_power}) after making the identification $\Delta_\mathcal{R}^2=1/(4 \pi^2 N^2 f_0)$.   Note that in previous treatments \cite{McFadden:2009fg, McFadden:2010na, McFadden:2010jw} we chose to Taylor expand this expression; here, we retain the full form to provide better accuracy in the case that $gq_*/q$ is not so small.

\begin{figure}[t]
\center
\includegraphics[width=4.5in]{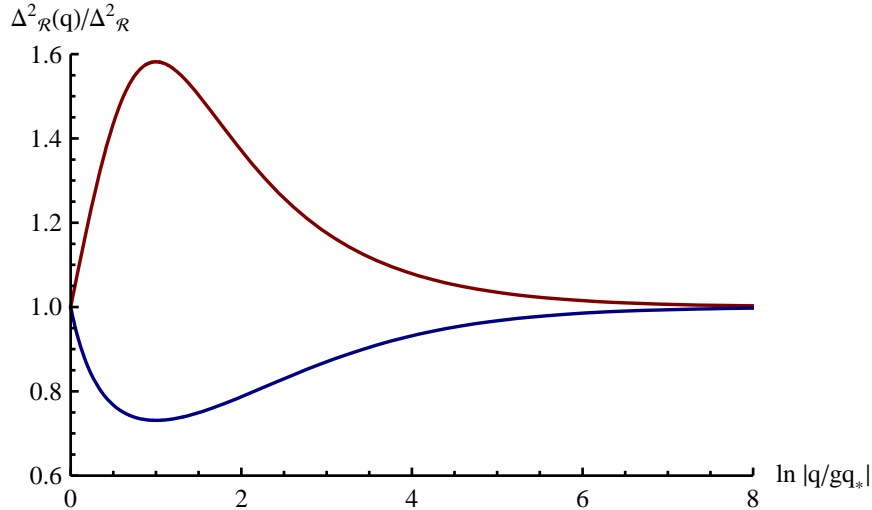}
\caption{\label{fig3}
Perturbative theoretical prediction for the power spectrum of the holographic model.  The lower curve corresponds to $g>0$ while the upper corresponds to $g<0$. The perturbative calculation is reliable for $\geff^2 \sim gq_*/q \ll 1$, corresponding to large momenta $q/gq_* \gg 1$ far from the peak/trough feature at $\ln|q/gq_*| = 1$.  At sufficiently high momenta, the power spectrum becomes nearly scale invariant, with $g>0$ corresponding to a blue tilt and $g<0$ to a red tilt.
}
\end{figure}

The power spectrum \eqref{hol_power} is plotted in Fig.~\ref{fig3} for both positive and negative $g$.
At sufficiently large momenta the spectrum rapidly becomes nearly scale invariant, with positive values of $g$ resulting in a slight blue tilt and negative values of $g$ yielding a slight red tilt.  This behavior reflects the fact that the dual QFT becomes asymptotically free at high momenta, with the free theory itself corresponding to an exact Harrison-Zel'dovich spectrum.

At lower momenta, the existence of the non-perturbative infrared scale $\qIR$ becomes apparent, resulting in the peak/trough feature in the spectrum at $q=egq_*$.
Note, however, that the perturbative calculation of $f(\geff^2)$ underpinning the power spectrum \eqref{hol_power} breaks down when $\geff^2 \sim gq_*/q$ becomes of order unity (recalling that $f_1$ is a constant of order unity).  This means that the perturbative result \eqref{hol_power} becomes unreliable at low momenta close to the peak/trough feature in Fig.~\ref{fig3}.
Moreover, our approximation $\beta=1$ is no longer justified in this regime and one should retain $\beta$ as an independent parameter, see Section \ref{sec:fudge}.
Since the smallest momentum scale appearing in the CMB is of the order $10^{-4}\,\mathrm{Mpc}^{-1}$, if the power spectrum \eqref{hol_power} is to reliably fit the entire range of CMB scales, then we conclude that the maximum value of $g$ is restricted to be of the order $|g|_{\mathrm{max}} \sim 2\times 10^{-3}$.  We will use this rough estimate later in setting the prior for $g$.

Note that the holographic model describes the very early universe; the end of this period should be the beginning of hot big bang cosmology.
The asymptotic (late-time) metric obtained at the end of the holographic period satisfies Einstein's equations. This follows from the fact that the 
cosmologies we analyze are related to holographic RG flows via the cosmology/domain-wall correspondence \cite{Skenderis:2006jq},
and the corresponding domain-wall spacetimes are known to satisfy Einstein's equations near their conformal boundary
\cite{deHaro:2000xn, Kanitscheider:2008kd}. This then ensures 
the conservation of the standard gauge-invariant curvature perturbation $\zeta$, which is the field dual to the trace of the stress energy tensor $T$. One thus expects a
smooth transition to hot big bang cosmology, with the holographic period supplying the initial conditions for the subsequent evolution.

Nevertheless, it would be very interesting to develop a detailed theory for the transition period, the analogue of the reheating period in conventional
scenarios, and here we offer a few preliminary comments leaving a more detailed study for future work. 
In order to exit the holographic period we would need to modify the UV structure of the dual QFT (since the UV of the QFT corresponds to late times). This can be achieved by adding irrelevant operators to the QFT. At momenta far below the momentum scale $q_{UV}$ set by the lowest dimension irrelevant operator, the computation of the 
2-point function (and therefore of the power spectrum) is well approximated by the computation described above.
Thus, as long as $q_{UV}$ is much larger that the largest momentum scale seen by CMB,
$q_{UV} \gg 10^{-1}$ Mpc${}^{-1}$, the error made by omitting this period is very small.
In principle, one could compute the corrections to the theoretical formula due to such irrelevant operators
and extract from the data the best-fit value for $q_{UV}$. We leave such study for future work, but we note that the ability to fit the data well without these corrections indicates they are indeed small.\footnote{Note however that the fit of the holographic model to data 
is better at low $q$ than high $q$, see Section \ref{sec:mod_comp}.  Perhaps the fit
at high $q$ would improve by including $q_{UV}$ in the analysis.}


\section{Parameter estimation} \label{sec:param_est}

Let us now turn to a comparison of the model with the data.  Holographic $\Lambda$CDM is described by the following free parameters, ($\Omega_bh^2$, $\Omega_ch^2$, $\theta$, $\tau$, $\Delta_\mathcal{R}^2$, $g$). The first five are common with the usual parametrization of $\Lambda$CDM\footnote{Note the parameter $\theta$ is the ratio between the sound horizon at the time of last scattering and the angular diameter distance of the surface of last scattering. Physically, this ratio fixes  the position of the acoustic peaks and  $\theta$ is tightly constrained by the data.  Theoretically, $\theta$ is a function of $\Omega_bh^2$, $\Omega_ch^2$ and $h$, so given $\Omega_bh^2$ and $\Omega_ch^2$, it may be expressed in terms of the dimensionless Hubble parameter $h$. For more details see, {\it e.g.}, Section 7.2 of~\cite{Weinberg:2008zzc}. }; the sixth parameter is related to the coupling of the dual gauge theory, as discussed in the previous section. For now, we will take the power spectrum~\eqref{hol_power} at face value. In the following section we compare the model to $\Lambda$CDM and calculate the Bayesian evidence after restricting $g$ to values for which the assumptions used in deriving the power spectrum are satisfied.

We perform our analysis for three data sets: the seven-year WMAP data~\cite{Komatsu:2010fb}, and two combinations of data sets introduced in~\cite{Komatsu:2010fb} as WMAP+BAO+$H_0$ and WMAP+CMB. The former is a combination of WMAP7 with priors on the Hubble constant~\cite{Riess:2009pu} and angular diameter distances~\cite{Percival:2009xn}. The latter is a combination of WMAP7 with  small-scale CMB experiments.
As usual, to compute the posterior probability distribution on the parameter space of the model given the data, we use Bayes' theorem which, for a flat prior probability distribution, relates it to the likelihood, {\it i.e.}, the probability for the data given a choice of parameters of the model
\begin{equation}
P(\alpha_M|D)\propto P(D|\alpha_M)\equiv\mathcal{L}(\alpha_M)\,,
\end{equation}
where $\alpha_M$ collectively denotes the parameters of the model $M$. For a given point in parameter space the likelihood can easily be evaluated, but the high dimensionality of the parameter space makes a grid-based approach prohibitive and other methods are required to sample it.
We will present results from an analysis using CosmoMC~\cite{Lewis:2002ah,cosmomc} with the MultiNest sampler~\cite{Feroz:2007kg,Feroz:2008xx}. We have also performed a more traditional Markov chain Monte Carlo analysis for the WMAP7 data set and found excellent agreement between the two. Both analyses made use of CAMB~\cite{Lewis:1999bs,camb} to evaluate the angular power spectra, version 1.5 of recfast~\cite{Seager:1999km} to compute the reionization history, and version 4.1 of the WMAP likelihood code~\cite{lambda} to evaluate the likelihood.

As a consistency check, we first performed a parameter estimation for $\Lambda$CDM and found good agreement with the values determined by the WMAP team.\footnote{The slight but consistent increase in the value of the scalar
spectral index can be traced to changes between recfast version 1.5 used in
this work and recfast version 1.4 used in the WMAP7 analysis.} Since we used a pivot scale $q_*=0.05\,\text{Mpc}^{-1}$, we repeat the parameters in Table~\ref{tab:parestlcdm} to facilitate comparison.
\begin{table}[t]
\begin{center}
\begin{tabular}{|l|c|c|c|}\hline
                           &WMAP7           &WMAP+BAO+$H_0$     &WMAP+CMB   \\\hline\hline
$\Omega_bh^2$              & $0.02252\pm0.00056$&$0.02257\pm0.00053$ &$0.02265\pm0.00051$\\\hline
$\Omega_ch^2$              & $0.1116\pm0.0054$&$0.1127\pm0.0035$ & $0.1124\pm0.0048$\\\hline
$100 \theta   $            &$1.0394\pm0.0027$& $1.0400\pm0.0026$&$1.0411\pm0.0022$\\\hline
$\tau   $                  &$0.088\pm0.014$& $0.088\pm0.014$& $0.088\pm0.014$ \\\hline
$\Delta_\mathcal{R}^2(q_*)$&$(2.183\pm0.073)\times10^{-9}$ &$(2.191\pm0.075)\times10^{-9}$  & $(2.190\pm0.068)\times10^{-9}$\\\hline
$n_s   $                     &$0.969\pm 0.014$&$0.970\pm 0.012$ &  $0.969\pm0.013$\\\hline
\end{tabular}\caption{\label{tab:parestlcdm} This table shows a summary of the parameters of the $\Lambda$CDM model and their uncertainties at the 68\% confidence level. }

\mbox{}

\begin{tabular}{|l|c|c|c|}\hline
                           &WMAP7           &WMAP+BAO+$H_0$     &WMAP+CMB   \\\hline\hline
$\Omega_bh^2$              & $0.02310\pm0.00045$&$0.02312\pm0.00043$ &$0.02326\pm0.00045$\\\hline
$\Omega_ch^2$              & $0.1077\pm0.0051$&$0.1120\pm0.0036$ & $0.1076\pm0.0042$\\\hline
$100 \theta   $            &$1.0407\pm0.0026$& $1.0406\pm0.0026$&$1.0423\pm0.0022$\\\hline
$\tau   $                  &$0.087\pm0.015$& $0.084\pm0.015$& $0.088\pm0.016$ \\\hline
$\Delta_\mathcal{R}^2$&$(2.146\pm0.088)\times10^{-9}$ &$(2.172\pm0.086)\times10^{-9}$  & $(2.151\pm0.084)\times10^{-9}$\\\hline
$g   $                     &$-0.00127\pm 0.00093$&$-0.00136\pm 0.00094$ &  $-0.00114\pm0.00088$\\\hline
\end{tabular}\caption{ \label{tab:parest}  This table shows a summary of the parameters of the holographic model and their uncertainties at the 68\% confidence level.}
\end{center}
\end{table}

The results for the mean values of the marginalized posterior distributions of the six parameters of the model and their uncertainties at $68\%$ confidence level are summarized in Table~\ref{tab:parest}.
All of them are for a pivot scale $q_*=0.05\,\mathrm{Mpc}^{-1}$, but notice that changing the pivot scale has only the trivial effect of rescaling the parameter $g$. Notice that for the best-fit values for $g$, the condition for parametric control $gq_*/q\ll1$ no longer holds for the lowest CMB momenta,
and the data appear to push us to intermediate values of the coupling. We will return to this issue in Section~\ref{sec:fudge}.

\begin{figure}
\begin{center}
\includegraphics[width=6.5in]{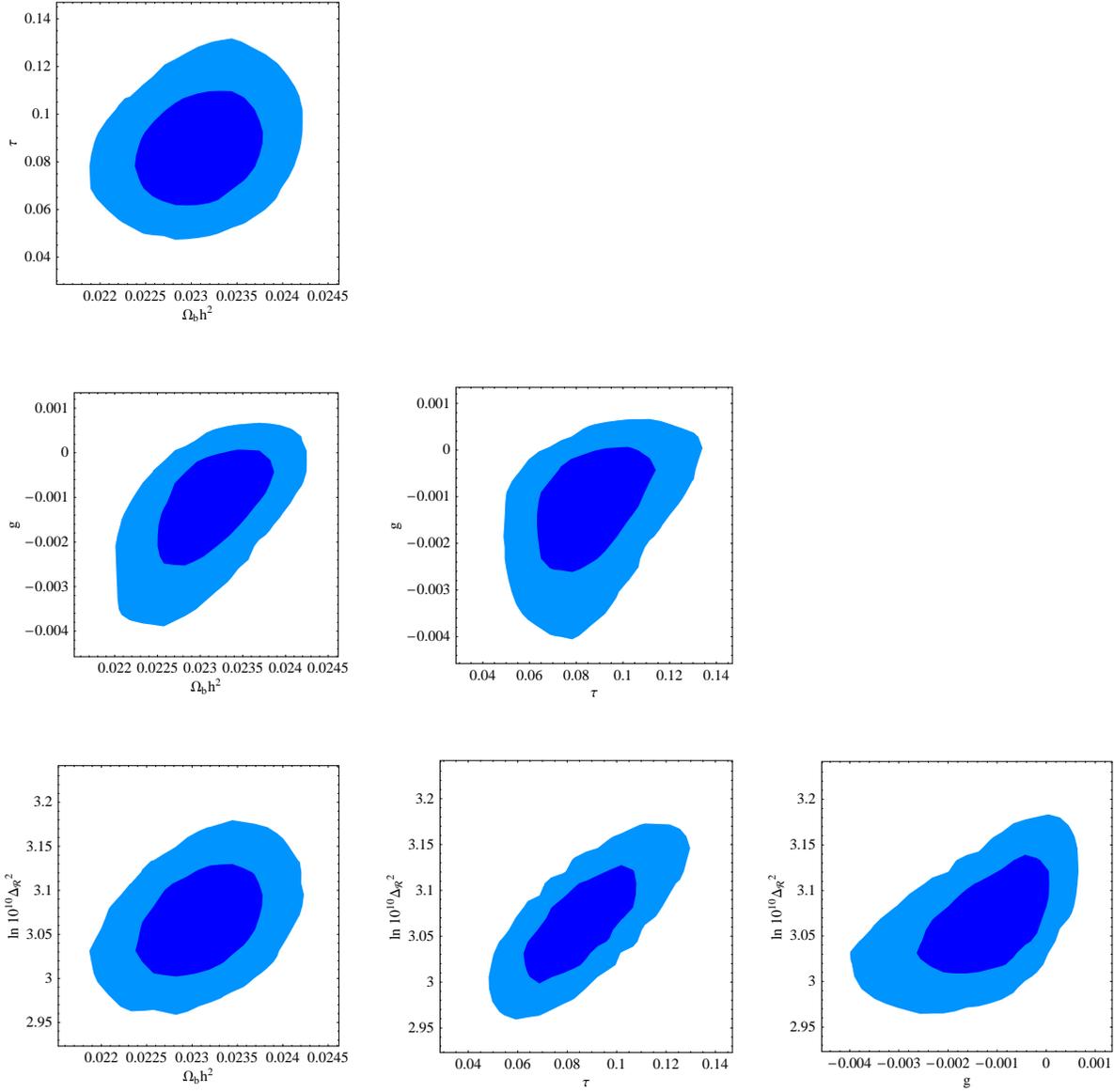}
\caption{A triangle plot showing the parameters of the holographic model derived from WMAP7. The contours represent 68\% and 95\% confidence levels.}
\label{markovwmap7}
\end{center}
\end{figure}

\begin{figure}[t]
\begin{center}
\includegraphics[width=6.5in]{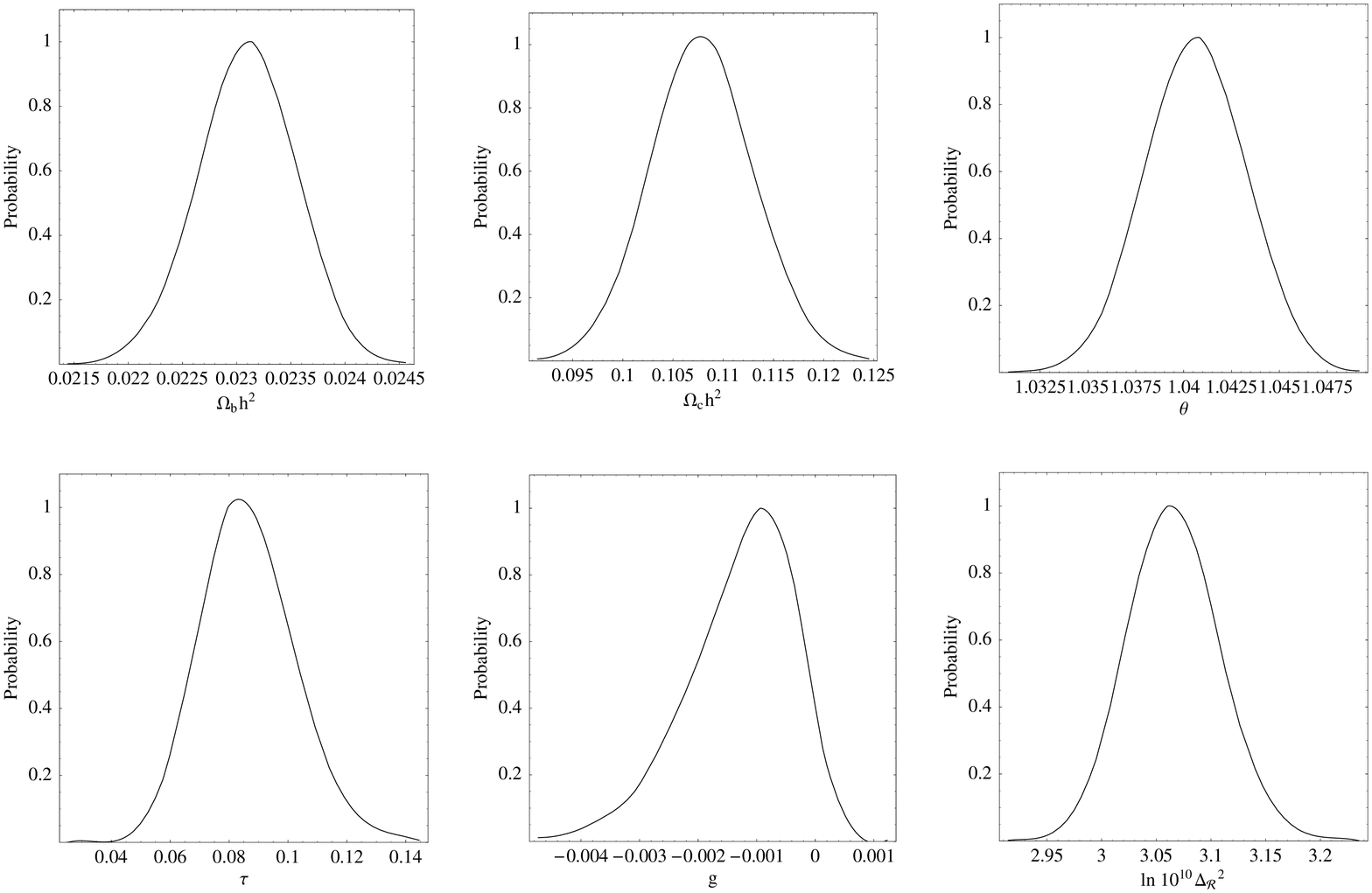}
\caption{The marginalized likelihoods for the parameters of the holographic model as derived from WMAP7. }
\label{singleparam}
\end{center}

\end{figure}

The estimated values of the parameters shared between the holographic and conventional $\Lambda$CDM models all overlap in both scenarios. However, we see that the central value of $\Omega_b h^2$ has moved by around one standard deviation.
There are some degeneracies between the new parameters in the power spectrum and the parameters characterizing the background geometry.
In Fig.~\ref{markovwmap7}, we display these degeneracies in the form of a triangle plot for the WMAP7 analysis, and  give the results for the individual variables in Fig.~\ref{singleparam}.

\section{Model comparison} \label{sec:mod_comp}
We now turn to a comparison between the holographic model and $\Lambda$CDM. Two approaches
have commonly been used to select models in the cosmology literature. The first
is based on information criteria; these typically
reward a larger best-fit likelihood
while penalizing the introduction of extra parameters. The second approach,
based on the Bayesian evidence,
rewards the model that provides the best fit when averaged over its parameter space. Information criteria
are relatively easy to evaluate since we already know the value of the likelihood for the best-fit point from the parameter estimation.
The Bayesian evidence is harder to evaluate because it involves an integral of the likelihood function over the parameter space of the model, but it allows constraints on the parameters to be taken into account more elegantly.
This is especially important in the case of the holographic model since
the form of the power spectrum \eqref{hol_power} is only valid when the gauge theory is in the perturbative regime, hence
there are natural priors that should be imposed on its parameters. We will discuss this in more detail in Section~\ref{subsec:Bayes}; we begin with a naive comparison using the best-fit likelihood.

\subsection{A first look: information criteria}
The most commonly used information criteria are the Akaike information criterion~\cite{akaike}
\begin{equation}
\text{AIC}=-2\ln\mathcal{L}_\text{best}+2k\,,
\end{equation}
where $k$ is the number of parameters characterizing the model, and the Bayesian or Schwarz information criterion~\cite{schwarz78}
\begin{equation}
\text{BIC}=-2\ln\mathcal{L}_\text{best}+k \ln N\,,
\end{equation}
where $N$ is the number of data points. Since the holographic model in the regime of small coupling and the $\Lambda$CDM model possess the same number of parameters, both criteria reduce to a comparison of the value of the logarithm of the likelihood function at the best-fit point. We summarize the best-fit likelihoods for the two models in Table~\ref{tab:bestlike}.
\begin{table}[t]
\begin{center}
\begin{tabular}{|l|c|c|c|}\hline
                   &Holographic Model & $\Lambda$CDM& $\Delta\ln\mathcal{L}_\text{best}$\\\hline\hline
WMAP7              &3735.5           &3734.3    & 1.2\\\hline
WMAP+BAO+$H_0$     &3737.3           &3735.7    & 1.6\\\hline
WMAP+CMB           &3815.0           &3812.5    & 2.5\\\hline
\end{tabular}\caption{ \label{tab:bestlike}  This table summarizes the best-fit values for ${-}\ln\mathcal{L}$ for both the holographic model and $\Lambda$CDM, as well as the difference between them. Positive numbers in the last column favor $\Lambda$CDM. The errors on our best-fit log likelihoods are estimated to be around $0.1$.}
\end{center}
\end{table}

From Table~\ref{tab:bestlike}, we may infer for example that, with all parameters tuned to their best-fit values, obtaining the observed WMAP data  is approximately three times more probable according to $\Lambda$CDM than according to the holographic model.
Note, however, that the probability of obtaining the data given the model with specific parameter values is {\it not} the same as the probability of the model given the data; it is the latter quantity that we really wish to evaluate and to which we now turn our attention.

\subsection{A closer look: Bayesian evidence}\label{subsec:Bayes}
Before presenting our results, let us quickly review Bayesian evidence. For a more detailed explanation, we refer the reader to~\cite{JaynesBK,Trotta:2008qt, LiddleBook} and references therein.

Our goal will be to compare the holographic model and $\Lambda$CDM, but we will keep our discussion general for now.  A given model is labelled by  $M$, and  characterized by a set of parameters $\alpha_M$.  In general, some of the parameters may overlap for different models, while others may be different. In the previous section, we determined the values $\alpha_M$ for model $M$ that are most likely given the data, {\it i.e.}, maximizes $P(\alpha_M|D)$. We now ask which {\em model\/} is most likely given the data, {\it i.e.}, which choice of  $M$ maximizes $P(M|D)$. Again we  use Bayes' theorem to relate this to the probability for the data given the model,
\begin{equation}
P(M|D)=\frac{P(D|M)P(M)}{P(D)}\,.
\end{equation}
The unconditional probability for the data appearing in the denominator is a model-independent constant and may be eliminated in the comparison of two models by taking the ratio
\begin{equation}
\frac{P(M_1|D)}{P(M_2|D)}=\frac{P(D|M_1)}{P(D|M_2)}\frac{P(M_1)}{P(M_2)}\,.
\end{equation}
This is sometimes referred to as the posterior odds. The first factor on the right-hand side is often called the Bayes factor, while the second factor is the ratio of prior probabilities for the models under comparison.
Typically, one assumes that all models are equally likely and this factor is set to unity.  The quantity of interest is $P(D|M)$, which is called the {\it evidence} and often denoted $E$. The probability for the data given a certain model is of course nothing but the probability for the data given a certain choice of parameters, $P(D|\alpha_M)=\mathcal{L}(\alpha_M)$, times the probability that this choice of parameters is realized, $P(\alpha_M)$, integrated over the space of parameters of the model. Consequently,
\begin{equation}
E=\int d\alpha_M P(\alpha_M)\mathcal{L}(\alpha_M)\,.
\end{equation}
Numerically, the evidence is difficult to evaluate because it  is effectively a multidimensional integral over   a function which is computationally nontrivial.  However, an efficient solution to this problem was found by Skilling~\cite{Skilling}, and is implemented within  MultiNest~\cite{Feroz:2007kg,Feroz:2008xx}.

As in parameter estimation, we need a prior probability distribution for the parameters $\alpha_M$. We will work with flat priors: namely, a prior probability that is constant over some defined region, and zero outside it. In this case, the evidence reduces to the integral
\begin{equation}
E=\frac{1}{\text{Vol}_M}\int d\alpha_M \mathcal{L}(\alpha_M)\, ,
\end{equation}
where $\text{Vol}_M$ is the volume of the region in parameter space over which the prior probability distribution is non-zero.  If we have a strongly peaked likelihood function which has support over only a relatively small region inside  $\text{Vol}_M$,  changing the  prior region can strongly affect the computed   evidence.  Provided the changes to the overall volume of the parameter space do not add or exclude regions where $\mathcal{L}$ is large, the integral will be unaffected while  $\text{Vol}_M$  can change substantially and the computed evidence is inversely proportion to $\text{Vol}_M$.

With the exception of $n_s$ and $g$ both models have the same
parameters. By using the same priors for the variables shared by the
holographic and standard $\Lambda$CDM scenarios the ambiguity in the
evidence associated with $\text{Vol}_M$ is minimized.  However, the
situation with $g$ and $n_s$ is more nuanced.  For the holographic
model we should restrict $g$ to values where perturbative expansion
used to derive \eqref{hol_power} is valid.  There are several unknown
factors of order unity in this perturbation expansion, so the
permissible range of $g$ is not specified precisely, but we therefore
simply estimate the largest allowed value of the coupling
$|g|_\text{max}$, permitting values of either sign.  As discussed in
Section \ref{sec:theo-pre}, we determine the value of $|g|$ for which
the naive expansion parameter $|g|q_*/q$ becomes unity at the longest
scales to which the CMB is sensitive, yielding $|g|_\text{max}\approx 2\times 10^{-3}$.
We then check whether the evidence depends strongly
on this choice by setting $|g|q_*/q$ to values between $1$ and $0.005$
at the longest scales, the latter choice giving $|g|_\text{max}\approx
10^{-5}$ and find that the evidence is only mildly dependent on the
choice of $|g|_\text{max}$.

The prior for $n_s$ is less obvious since, unlike $|g|_\text{max}$, it
is a purely empirical parameter and we cannot restrict it by appealing to
the internal consistency of some underlying theory.  Moreover, our
best information about $n_s$ is derived from the WMAP data we are
using to compute the evidence, and it would be inappropriately
circular to set the prior on $n_s$ directly from a parameter estimate
derived from the WMAP data itself.  On the other hand, choosing an
overly generous range for $n_s$ reduces the evidence and may
overestimate the ability of competing models to explain the data.  To
account for this dilemma, we work with two priors for $n_s$,
$0.92<n_s<1.0$ and $0.9<n_s<1.1$.  As is well known, the first choice
includes essentially the entire range over which the likelihood is
appreciably different from zero while the second is centered on the
scale invariant Harrison-Zel'dovich spectrum.  For these choices the
ratio of $\text{Vol}_M$ is $0.08/0.2=0.4$, shifting $\ln E$ by about
$0.9$.  In order to conclusively favor one model over another we would
require a much larger difference in the respective values of the
evidence, so this ambiguity will not qualitatively alter our
results. For completeness, we summarize the priors on all the
parameters of $\Lambda$CDM in Table 4. 

\begin{table}[tb]
\begin{center}
\begin{tabular}{|l|c|c|}\hline
   $ X$                    &$X_\text{min}$ & $X_\text{max}$ \\\hline\hline
$\Omega_bh^2$              &$0.020 $&$0.025$\\\hline
$\Omega_ch^2$              &$0.09$ & $1.25$\\\hline
$100 \theta   $            &$1.03$ &$1.04$\\\hline
$\tau   $                  &$0.02$ &$0.15$\\\hline
$\Delta_\mathcal{R}^2(q_*)$&$1.82\times 10^{-9}$ & $2.71\times 10^{-9}$\\\hline
$n_s$ ~ (1)                   &$0.92$ &$1.0$\\\hline
$n_s$ ~ (2)                   &$0.9$ &$1.1$\\\hline
\end{tabular}\label{tab:priorlcdm}\caption{This table shows the priors used in the calculation of the evidence for the $\Lambda$CDM model. The two $n_s$ priors correspond to the narrow and broad choices discussed in the text.}
\end{center}
\end{table}

We also  consider  asymmetric priors in which the coupling $g$ ranges from zero to a value of either $+|g|_\text{max}$ or $-|g|_\text{max}$. We present results  for $|g|_\text{max} = 1\times 10^{-4}$ and also $|g|_\text{max} =  2\times 10^{-3}$. The latter value is rather large, so we use it with the caveat that the assumptions implicit in our use of the holographic power spectrum may begin to break down for momenta that contribute to the lowest multipoles in the angular power spectrum.  However, if we assume that the power spectrum can be trusted in this scenario we find moderate evidence suggesting $g<0$ using present data.  Consequently, it seems worth exploring both the reliability of (\ref{hol_power}) at higher values of $|g|_\text{max}$ and attempting to determine whether fundamental theory predicts the sign of $g$, as answers to these questions could significantly improve our ability to constrain holographic $\Lambda$CDM, even with present-day data.

\begin{figure}[t]
\begin{center}
\includegraphics[width=6.0in]{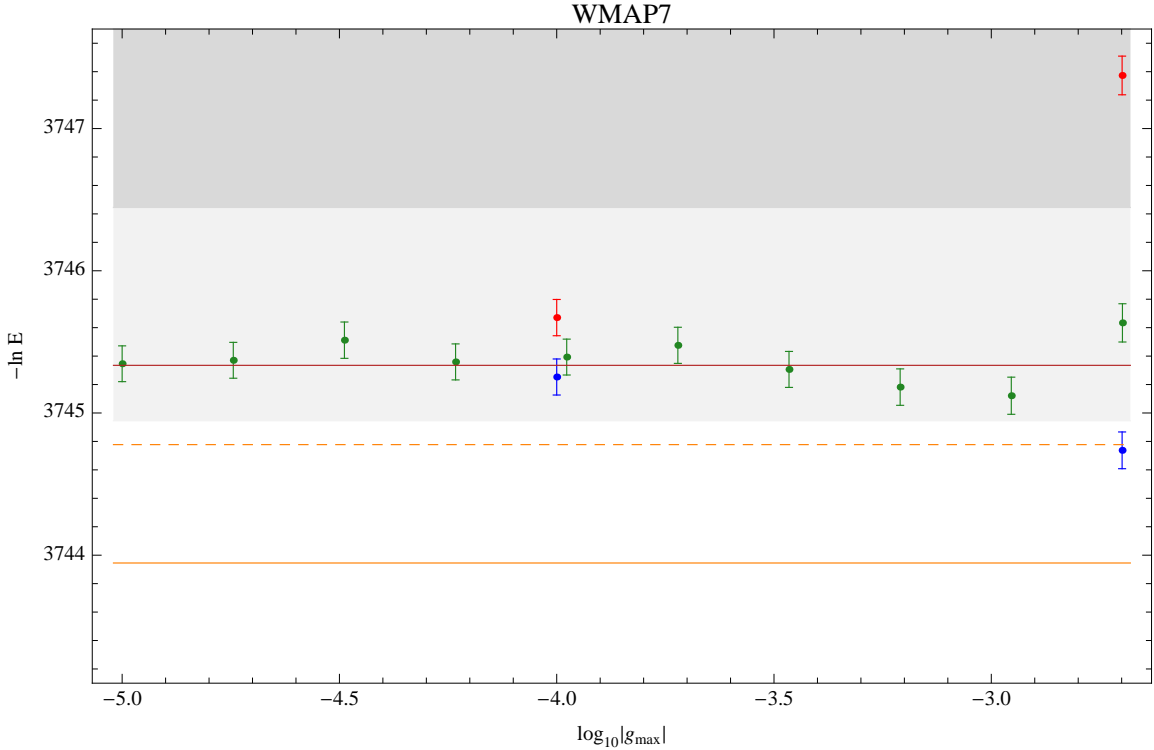}
\caption{This figure shows the results of our calculation of the evidence for the WMAP7 data set. $\Lambda$CDM is shown as a solid orange line and a dashed orange line for the narrow and broad priors on $n_s$ discussed in the text, respectively. The red line is the evidence for the pure Harrison-Zel'dovich spectrum.  The green data points represent evidence computed for the holographic model, as a function of  $|g|_\text{max}$, as indicated on the horizontal axis. The red and blue points show the holographic model with asymmetric priors with positive and negative $g$, respectively. The shading indicates the difference in evidence according to the Jeffreys scale (see end of Section \ref{subsec:Bayes}), relative to $\Lambda$CDM with the narrow prior.  }
\label{evwmap7}
\end{center}
\end{figure}

\begin{figure}[t]
\begin{center}
\includegraphics[width=6.0in]{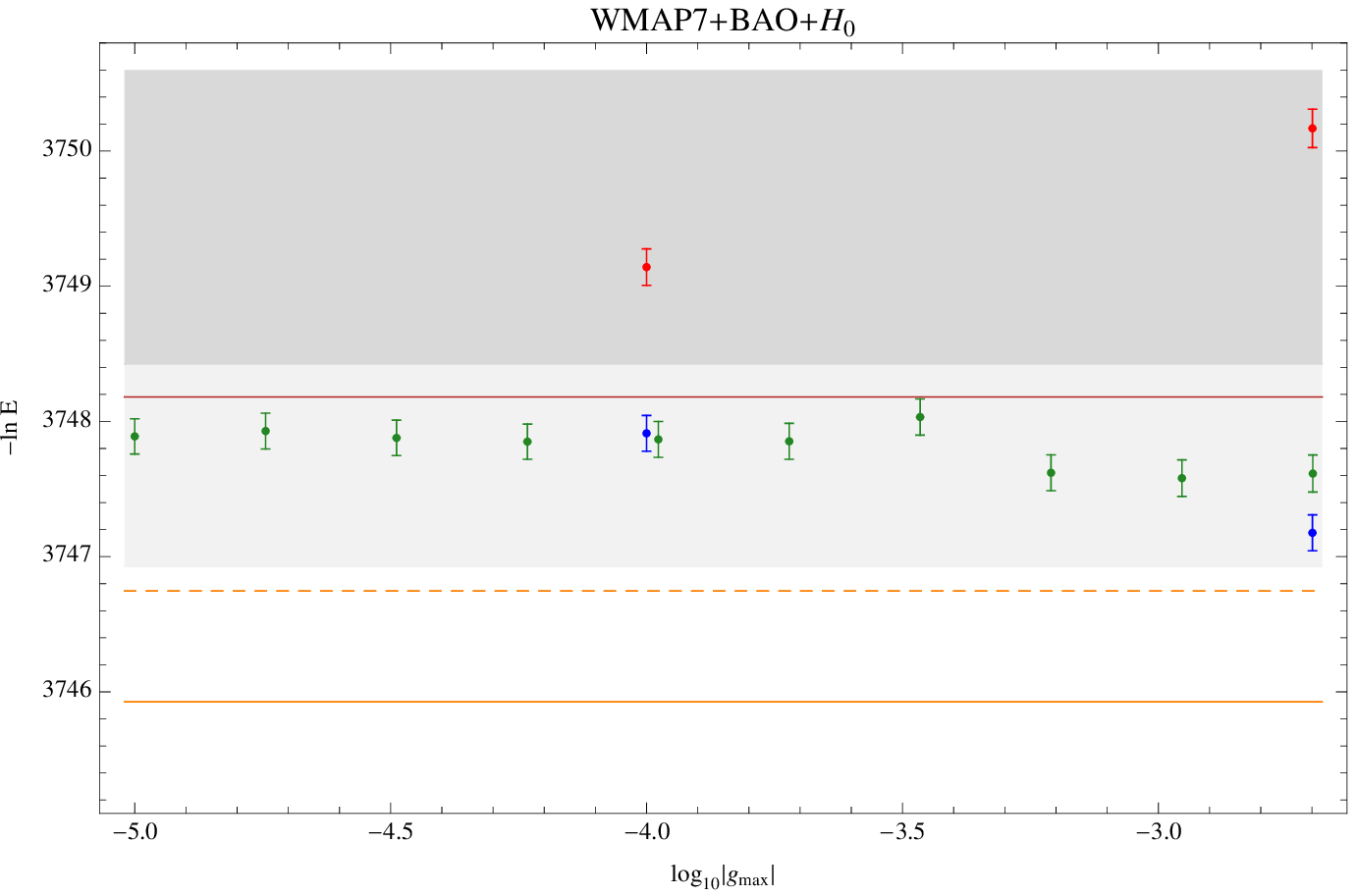}
\caption{This figure shows the results of our calculation of the evidence for the WMAP+BAO+$H_0$ data set, with the same conventions as Fig.~\ref{evwmap7}.  }
\label{evbao}
\end{center}
\end{figure}

\begin{figure}[t]
\begin{center}
\includegraphics[width=6.0in]{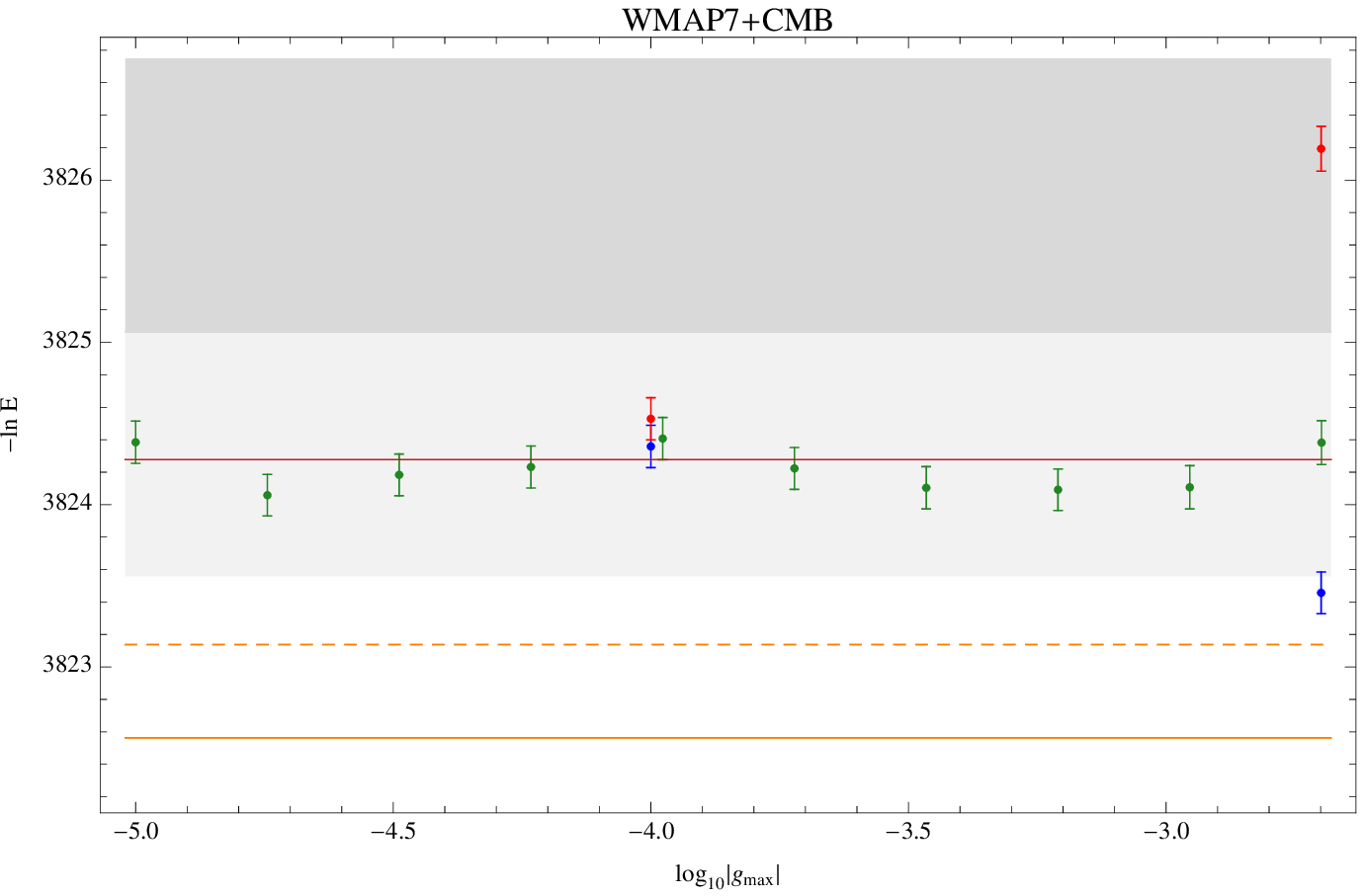}
\caption{This figure shows the results of our calculation of the evidence for the WMAP+CMB data set, with the same conventions as Fig.~\ref{evwmap7}.     }
\label{evcmb}
\end{center}
\end{figure}

We show the evidence computed using  WMAP7 on its own and the combinations  WMAP\ +BAO\ +$H_0$ and   WMAP+CMB   in Figs.~\ref{evwmap7},~\ref{evbao} and~\ref{evcmb}, respectively. Along with the evidence for $\Lambda$CDM and the holographic model, the plots also show the evidence for the Harrison-Zel'dovich spectrum. As a guide to the eye, we  shade the plots to indicate regions with $\Delta\ln E<1$ (white), $1<\Delta\ln E<2.5$ (light gray), and $2.5<\Delta\ln E<5$ (darker gray).  In the terminology of the Jeffreys scale, these intervals are ``not worth more than a bare mention'', ``significant'', and ``strongly significant'' \cite{Trotta:2008qt}.  On the basis of the narrower prior $0.92<n_s<1.0$, we find that there is only weak to moderate evidence in favor of $\Lambda$CDM based on the data sets we have studied.

When $g$ is restricted to small values, the evidence we compute for holographic $\Lambda$CDM approaches the value found for the scale-free Harrison-Zel'dovich spectrum. This is easily understood: in this limit, the power spectrum (\ref{hol_power}) is only weakly dependent on values of $q$ which contribute to the CMB.  Consequently, $\Delta_\mathcal{R}^2(q)$ depends very weakly on $g$, and $\partial \mathcal{L}/\partial g$ is close to zero.  Thus the evidence integral effectively factorizes and the resulting $\int dg$ term in the numerator will cancel the corresponding term in $\text{Vol}_M$. In this limit the expression for the evidence reduces precisely to  the Harrison-Zel'dovich variant of $\Lambda$CDM.

\subsection{Sensitivity to the infrared scale}  \label{sec:fudge}

As we discussed in Section \ref{sec:theo-pre}, the 2-loop QFT
calculation for $f(\geff^2)$ is sensitive to an infrared scale $\qIR
\sim \gYM^2N$. This is reflected in the presence of the $f_2$ term in
(\ref{f}), or equivalently the factor $f_3$ in (\ref{f_expansion}).  As
long as one probes the theory on scales far above $\qIR$, for which
$\geff^2$ is very small, we have $|\ln gq_*/q|\gg |\ln f_3|$. These
terms therefore make only a very small contribution and their precise numerical
values are of no significance.

In our case, the scales that we probe are fixed:
the WMAP momentum range is approximately
$10^{-4}\lesssim q \lesssim 10^{-1}\,\mathrm{Mpc}^{-1}$
but the size of $\gYM^2$, or equivalently $g$, is unknown.
For sufficiently small $g$, the
infrared scale $\qIR$ is outside the WMAP momentum range
and the theory is then insensitive to value of $f_3$
(since $|\ln gq_*/q|\gg |\ln f_3|$ for all modes in the CMB).
This has been our reasoning in setting $f_3=|f_1|$ thus far.

For larger values of $g$, however, the infrared scale $\qIR$ is
inside the WMAP momentum range and the long wavelength modes in the CMB
are  sensitive to the infrared scale. In this case,  it is not a good approximation to set $f_3=|f_1|$, but we we should take $f_3=\beta|f_1|$, where the infrared parameter $\beta$ encodes the location of the non-perturbative scale $\qIR$.  In principle, $\beta$ is fully determined by the dual QFT and is not an
additional parameter of the theory.   In practice, however,  computing
$\beta$ is not an easy task and requires non-perturbative information
({\it e.g.}, a large-$N$ resummation, see \cite{Jackiw:1980kv,Tom}).
However, we can determine the value of the coupling for which the
data becomes sensitive to  $\beta$. For very small values of $g$, the power
spectrum~\eqref{hol_power} used in the 
previous sections accurately reflects the predictions of the
holographic model and is effectively independent of $\beta$, so its precise values does not matter.  However, there will be a range of values of the coupling for which both the spectrum is sensitive to the specific value of $\beta$ and the perturbative expansion is still valid, and in this case our spectrum (\ref{hol_power}) is incomplete.  We determine the value at which $\beta$ becomes important  by fitting to the power spectrum
\be
\label{fudgedpowerspec}
\Delta_\mathcal{R}^2(q) =\Delta_\mathcal{R}^2 \frac{1}{1+ (g q_*/q) \ln |q/\beta g q_*|}\,,
\ee
and computing the marginalized probability distribution for  $\beta$ with different values of $|g|_\text{max}$.

\begin{figure}[htp]
\begin{center}
\includegraphics[width=5.5in]{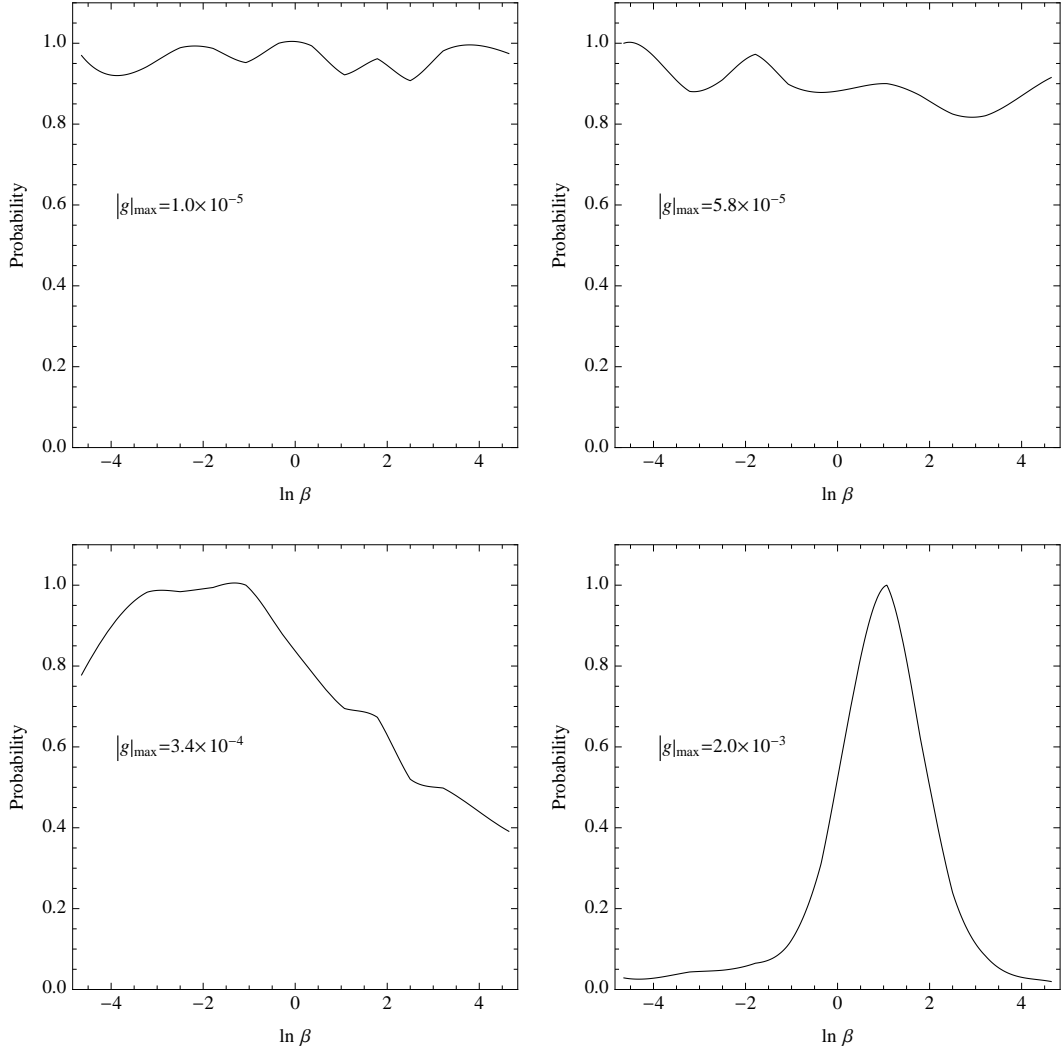}

\caption{This plot shows the marginalized probability distribution for the infrared parameter $\beta$ for various values of the coupling. The upper left corresponds to $|g|_\text{max}=0.002$, the upper right to $|g|_\text{max}=3.4\times 10^{-4}$, the lower left to $|g|_\text{max}=5.8\times 10^{-5}$, and the lower right to $|g|_\text{max}=10^{-5}$.  \label{fig:fudge}}
\end{center}

\end{figure}

The result is displayed in Fig.~\ref{fig:fudge}. As expected,  for $|g|_\text{max}\sim 10^{-5}$, the marginalized probability distribution for $\beta$ is approximately flat over several orders of magnitude, and $\beta$ is an irrelevant parameter. Raising $|g|_\text{max}$ to $2\times 10^{-3}$,  we see that the probability distribution for $\beta$ is peaked near $\beta\sim1$.  For couplings $|g|$ of this magnitude, we are quite sensitive to the presence of the infrared scale. It is conceivable that the correct value is indeed of order unity, but it seems equally likely that $\beta$  is a factor of ten larger or smaller than unity, in which case we are far from the result that would be returned by a precise calculation of the spectrum.  The transition is not very sharp and occurs around $|g|\sim10^{-4}$.
We have seen that the best-fit points for the holographic model for the three data sets all lie at values of $|g|$ about an order of magnitude larger than this. It thus seems that the data may be pushing us into a regime of intermediate coupling where the predictions become infrared sensitive.   This motivates further theoretical work to establish the predictions of the holographic model in this regime.

\section{Conclusions}

In this paper, we used the WMAP data to test the holographic $\Lambda$CDM model, and we conclude that the holographic model is compatible with current data.  To compare the holographic model with conventional power-law $\Lambda$CDM, we evaluated  Bayesian evidence for both models, paying close attention to the choice of priors.    The prior on $g$, the coupling that controls the holographic model, is set to include only values for which the perturbation expansion underpinning the holographic power spectrum is  valid over the entire range of momenta relevant to the CMB.
For the power-law $\Lambda$CDM model,  $n_s$ is a purely empirical parameter so the corresponding prior cannot be deduced on the basis of a purely physical argument.  Consequently, we consider two different choices for prior; the first is a near optimal choice containing the entire range over which the likelihood is appreciably
different than zero and the second is centered on the scale invariant Harrison-Zel'dovich spectrum.
In both cases there is at most weak evidence for  $\Lambda$CDM, relative to the holographic scenario, although the difference in evidence grows somewhat more pronounced if ground/balloon-based CMB data is added to the fit, or if we include 
measurements of baryon acoustic oscillations and $H_0$.  More precisely, with the narrow prior  of $0.92<n_s<1$ for the $\Lambda$CDM spectral tilt
the difference in $-\ln E$ is of order $1.2$ to $1.6$, see Figs.~\ref{evwmap7}, \ref{evbao} and \ref{evcmb}, while if one assumes the
prior of $0.9<n_s<1.1$, the difference in evidence is no longer significant.  We conclude that at this point we do not find any strong evidence
in favor of $\Lambda$CDM.

We can also compare holographic $\Lambda$CDM to the exactly scale invariant Harrison-Zel'dovich power spectrum, and in this case the computed evidence for  both models is essentially identical, unless we allow the coupling $g$ to approach values
for which the underlying perturbation theory appears to become unreliable.
Note that the holographic model with $g=0$ ({\it i.e.}, free field theory in three dimensions)  predicts an exactly scale invariant Harrison-Zel'dovich power spectrum, and is the only known microphysical model to do so,  to our knowledge.  Thus, the current data are also consistent with the dual theory being
a free QFT.

Above all, the results of this paper call for an improved theoretical understanding of the predictions of the holographic model.
Our restriction on the maximum value of $|g|$, derived
from a naive estimate of when the effective coupling $\geff^2$ becomes of order unity for the lowest CMB scales, appears to
limit the holographic model to spectra with insufficient scale dependence.
Besides accounting for the similar performance of the holographic model and the Harrison-Zel'dovich spectrum,
this is consistent with the location of our best-fit estimate for $|g|$ near the top of the currently allowed range.
A precise determination of the relationship between $g$ and $\geff^2$, and hence of the maximum allowed value of $|g|$, is therefore important.
Specifically, this relationship is given by $f_1 \geff^2 = g q_*/q$, motivating a full 2-loop QFT calculation
of the constant $f_1$.
The result will in general depend
on the Yukawa and quartic couplings, as well as the field content of the theory.
Larger values for $f_1$ would permit a larger upper bound for the coupling $|g|$, allowing power spectra with a stronger scale dependence.
For example, if $f_1 \sim 10$, then $f_1 \geff^2 \sim 1$
implies $\geff^2 \sim O(10^{-1})$, hence the three and higher-loop contributions are (relatively) small, even though the 2-loop contribution is of order one.
On the other hand, if $\geff^2$ is not small for all relevant momenta, the higher order terms would be needed explicitly.

Even if the 2-loop approximation is sufficient, there is another related issue that must be addressed.
As we discussed in Section \ref{sec:theo-pre}, the complete 2-loop results are sensitive to an infrared scale $\qIR \sim \gYM^2 N$, leading to the $f_2$ term in (\ref{f}), or the $f_3$ factor in (\ref{f_expansion}).  On scales far above $\qIR$, these terms make only a very small contribution.  The range of momenta that contribute to the CMB are fixed,
so equivalently infrared effects are unimportant only when the coupling $g$ is sufficiently small.
To confirm this quantitatively, in Section \ref{sec:fudge} we parametrized the infrared effects via an undetermined constant $\beta$.
For  $|g|$ less than $\sim 10^{-4}$ the data are  insensitive to the precise value of $\beta$. However, this threshold is below the best-fit value $|g| \sim 10^{-3}$, so developing a better understanding of infrared effects is clearly important.
In this regard, a direct computation of $\beta$, either through large-$N$ resummation methods or through lattice simulations, 
appears the most promising line of enquiry.

On the observational side,   parameter estimates can be expected to tighten dramatically in the near future. In particular, the  Planck satellite will measure the power spectrum accurately over a wide range of angular scales,  with excellent signal to noise.  Assuming  broken scale-invariance at a similar level to the central values seen by WMAP, the Harrison-Zel'dovich limit of the holographic model will be ruled out with a very high degree of confidence.  Empirically, WMAP measures 
the running $\alpha_s$ in the usual running-$\Lambda$CDM model with roughly the same degree of precision as it measures the breaking of scale-invariance,  $n_s-1$ \cite{Komatsu:2010fb}.  The holographic model predicts that $\alpha_s \sim |n_s-1|$, so  it is perhaps unsurprising that the WMAP dataset does not permit us to make  a strong distinction between the holographic model and $\Lambda$CDM without running.
However, Planck will measure $\alpha_s$ at a much higher level of precision than the current central value of $n_s-1$, and future experiments will do even better \cite{Adshead:2010mc}. Consequently,  it appears that Planck will be able to rule out the holographic scenario discussed here if a running $\alpha_s\sim |n_s-1|$ is not detected \cite{McFadden:2010jw}.  In the event that such a running {\it is} detected, in order to distinguish the holographic model from running $\Lambda$CDM and other models permitting a strong running (for example, slow roll inflation with modulations \cite{Kobayashi:2010pz}), it will then be necessary to examine higher logarithmic derivatives of the power spectrum.  
The ability of Planck to discriminate between such models is less clear and merits further investigation.

One issue we have not addressed in detail here is the tensor power spectrum.  The ratio $r$ of tensor to scalar amplitudes predicted by the holographic model may be found in \cite{McFadden:2009fg, McFadden:2010na}: to leading order in $\geff^2$, $r$ is a constant depending on the field content of the dual QFT.  In particular, we emphasize that $r$ is not parametrically suppressed, in contrast to models of slow roll inflation.
We do not present detailed parameter fits including tensors, but using the WMAP dataset on its own with a tensor contribution  proportional to $\Delta_\mathcal{R}^2(q)$ we find a weak preference for a nontrivial spectrum of primordial gravitational waves, consistent with the results of \cite{Dias:2011in}.   
This is understandable, since tensor modes contribute to the temperature anisotropies at large angular scales and combining these with a nearly scale invariant spectrum of scalar perturbations  results in a better fit to the WMAP  data. The estimated distribution for $r$ peaks at zero for a fit to the WMAP7+H$_0$+BAO dataset, so we do not attach any particular significance to this result.
A primordial gravitational wave background is degenerate with $g$, and including this contribution drives the maximum likelihood value of $g$ closer to zero.
Moreover, Planck (and a variety of forthcoming ground and balloon-based CMB experiments) will  break this degeneracy by tightly constraining the B-mode of the CMB polarization, in addition to the primordial temperature anisotropies.

If future data provides strong evidence for holographic $\Lambda$CDM this would constitute the first observational evidence for holography, given that  these models do not have a conventional realization in terms of weakly coupled gravity coupled to other fields.
On the other hand, one should emphasize that the scenario discussed is a specific realization of holographic inflation: even the conventional $\Lambda$CDM model, realized in the usual way as an
inflationary model, also has a holographic realization in terms of a strongly coupled QFT. Moreover, there are
other holographic models based on deformations of conformal field theories (the first class in the discussion of Section 2), and these would in general have different observational signatures. It would be interesting to
extract the predictions for these models and confront them with observational data.  In summary, these studies thus present a unique arena
where theoretical ideas about Planck scale physics can be tested
observationally: an exciting period lies ahead of us.


\section*{Acknowledgments}

The work of R.E. and R.F.~is supported in part by the National Science Foundation under Grant No.~NSF-PHY-0747868, and the Department of Energy under Grant No.~DE-FG02-92ER-40704.  P.M.~and K.S.~are supported by NWO, the Nederlands Organisatie voor Wetenschappelijke Onderzoek.  This work was supported in part by the facilities and staff of the Yale University Faculty of Arts and Sciences High Performance Computing Center.


\end{document}